\documentclass[twocolumn,showpacs,preprintnumbers,amsmath,amssymb,prl]{revtex4}

\usepackage{graphicx}
\usepackage{dcolumn}
\usepackage{bm}

\begin{document}

\preprint{APS/123-QED}

\title{Experimental indication of anomalous sensitivity in
many-body systems:\\Deterministic randomness in complex quantum
collisions?}

\author{Qi Wang$^{1}$}
 \email{wangqi@impcas.ac.cn}
\author{Jian-Long Han$^{1,3}$}
\author{Yu-Chuan Dong$^{1}$}
\author{Song-Lin Li$^{1}$}
\author{Li-Min Duan$^{1}$}
\author{Hu-Shan Xu$^{1}$}
\author{Hua-Gen Xu$^{1}$}
\author{Ruo-Fu Chen$^{1}$}
\author{He-Yu Wu$^{1}$}
\author{Zhen Bai$^{1,3}$}
\author{Zhi-Chang Li$^{2}$}
\author{Xiu-Qin Lu$^{2}$}
\author{Kui Zhao$^{2}$}
\author{Jian-Cheng Liu$^{2}$}
\author{Guo-Ji Xu$^{2}$}
\affiliation{$^{1}$ Institute of Modern Physics, Chinese Academy
of Sciences, Lanzhou 730000, China\\
$^{2}$ China Institute of Atomic Energy, Beijing, 102413, China\\
$^{3}$ Graduate School of Chinese Academy of Sciences, Beijing
100049, China }

\author{S.Yu. Kun$^{4,5}$}
\email{kun@fis.unam.mx}
 \affiliation{
$^{4}$ Facultad de Ciencias, Universidad Auton\'oma del
Estado de Morelos (UAEM), 62209-Cuernavaca, Morelos, Mexico\\
$^{5}$Nonlinear Physics Center, RSPhysSE, The Australian National
University (ANU), Canberra ACT 0200, Australia
}%

\date{\today}

\begin{abstract}
We have experimentally tested a recently suggested possibility for
anomalous sensitivity of the cross sections of dissipative heavy ion collisions.
Cross sections for the $^{19}$F+$^{27}$Al dissipative collisions were 
measured at the fixed energy 118.75 MeV of the
$^{19}$F for the 12 different
 beam spots on the same target foil. The data demonstrate
dramatic differences between the cross
sections for the different beam spots. The effect may indicate
 deterministic randomness in complex quantum collisions.
New experiments are highly desirable in a view of the fundamental importance
of the problem.
\end{abstract}

\maketitle

For classical chaotic systems infinitesimally small uncertainties
in the initial conditions are exponentially enhanced after a
certain time making long time dynamics unpredictable.
 This unpredictability is referred to as
dynamical instability of motion or deterministic randomness~\cite{CasChir}. 
On the contrary, quantum systems with finite number of degrees of freedom, 
whose classical
counterparts are chaotic, are dynamically stable with respect to 
small changes of the initial state. Such a stability is also
 expected with respect to infinitesimally small
perturbations of the quantum systems. Indeed, effect of a small
perturbation on an excited quantum system is quantified by its
strength, $|K|$, as compared to the average level spacing $D$ of
the system. Here $K$ are matrix elements which couple unperturbed
states by the perturbation. From the perturbation theory, for
$|K|/D\ll 1$, the perturbation
 produces only a little effect on the system.

Alternative consideration~\cite{kunPRL} of the complex
collisions
has suggested an anomalous sensitivity of the cross sections to
the extremely small perturbation of the intermediate complex (IC)~\cite{App3}. In
particular, the perturbation with a strength
$|K|\sim 10^{-4}D$ or even smaller may result in $\pm 15\%$
variations of the cross sections of dissipative heavy ion
collisions (DHIC). The consideration~\cite{kunPRL} is intimately
related to, and provides an interpretation for, the
channel-channel correlation and nonself-averaging of the
excitation function oscillations in DHIC. A spontaneous character of setting up
the cross sections was supplemented by effect of the non-equilibrium phase
transitions for channel-channel correlations in DHIC~\cite{kunAIP}, which is
essentially a quantum interference phenomenon. 
In~\cite{kunPRL}, the small perturbation was
modeled by changing the number of electrons moving in Coulomb
field of the IC. However, if the effect of
the anomalous sensitivity does exist, this would imply other
intriguing possibilities. For example, one may experimentally study 
effects of chemical structure of the target and its phase, e.g.,
crystalline or amorphous, on the cross sections. 
Yet, before designing such
experiments, one has to experimentally address the problem of the
reproducibility of the cross sections for nominally identical
conditions in different experiments. 
This is because even nominally identical target foils have
different distributions of electromagnetic fields, defects, etc.
Such a non-reproducibility has indeed been revealed in the two
independent experiments~\cite{WQijmpe}.
However, because after the first measurement~\cite{WQijmpe}
the reaction chamber was dismantled it may not be excluded that
some experimental conditions could change for the second
measurement.

The fundamental importance of the problem has motivated us to test
the anomalous sensitivity of the cross sections in a new
experiment without opening the reaction chamber. 
Measurements of differential cross sections of 
fragments B, C, N and O produced in the DHIC
 $^{19}$F+$^{27}$Al have been carried out at the China
Institute of Atomic Energy (CIAE), Beijing. The beam of
$^{19}$F$^{9+}$ at incident energy 118.75 MeV was provided by the
HI-13 tandem accelerator, CIAE. Two sets of gas-solid ($\Delta
$E-E) telescopes, with a charge resolution Z/$\Delta $Z$\approx
$30 and an energy resolution $\approx $400 keV, were set at $\theta
_{lab}$=57$^\circ$ and 31$^\circ$. The angular acceptances of the
telescopes were $\delta\theta =17.5^\circ$ and $\delta\phi
=3.1^\circ$ for $\theta _{lab}$=57$^\circ$, and $\delta\theta
=6.4^\circ$ and $\delta\phi =1.3^\circ$ for $\theta
_{lab}$=31$^\circ$. The $\Delta $E detector is an ionization
chamber filled with a mixture-gas of 90$\%$ argon and 10$\%$
methane in flowing mode at a pressure of 100 mb. The residual
energy is deposited in a Si detector with a thickness of 500 $\mu
$m. The low energy thresholds of the $\Delta $E-E telescopes were
about 7,8,9 and 11 MeV for the  B, C, N and O fragments
respectively. Two silicon semiconductors arranged at $\theta
_{lab}=\pm 7.5^\circ$ were employed to monitor the beam, and a
Faraday cup was placed at $\theta _{lab}=0^\circ$. The $\Delta
$E-E telescopes, the monitors and a Faraday cup were placed in the
reaction plane. A 10$\times $50mm rectangular $^{27}$Al foil was
produced with vacuum evaporation method ~\cite{WQref2}. Its
average thickness of $\simeq$67 $\mu $g/cm$^{2}$ corresponds to 180
keV beam energy loss in the target. Before the experiment the
target thickness was measured with $\alpha$-particle thickness
gauge using the air equivalent method ~\cite{WQref2,WQref1} at 22
different target points with the step 2 mm along the straight line
on which 12 beam spots, in the following experiment, were located.
Relative variation of the thickness did not exceed 8$\%$.
 The total energy spread due to the
beam energy spread (45 keV) and the beam energy loss in the target
(180 keV) was about 225 keV, i.e. about a characteristic energy
length of the oscillations in the excitation functions for the
$^{19}$F+$^{27}$Al DHIC~\cite{RomCat}. Thus, fine energy
resolution in our experiment allows to resolve the energy
oscillating component of the cross section, which is suggested to
be responsible for the anomalous sensitivity of the cross sections~\cite{kunPRL}.

In the measurement, 12 target points were bombarded by moving the
rectangular target in steps of 2 mm along a fixed direction
perpendicular to the reaction plane. After the experiment the
target thickness was measured again at each of the 12 beam spots
having a diameter of about 1 mm which is close to the beam
diameter on the target.

At the beginning of the experiment we have measured angular
distributions of the
 B, C, N and O fragments at the E$_{lab}$=114 MeV incident beam energy (Fig. 1).
The angular distributions were obtained by integrating over the
whole outgoing spectra. As an example, in Fig. 1 we present 
energy spectrum for the N products at $\theta_{lab}=31^\circ$
obtained for one of the 12 beam spots. The angular
distributions and the spectrum show characteristic features of
DHIC~\cite{RomCat}.

\begin{figure}[!htbp]
\includegraphics[width=8cm]{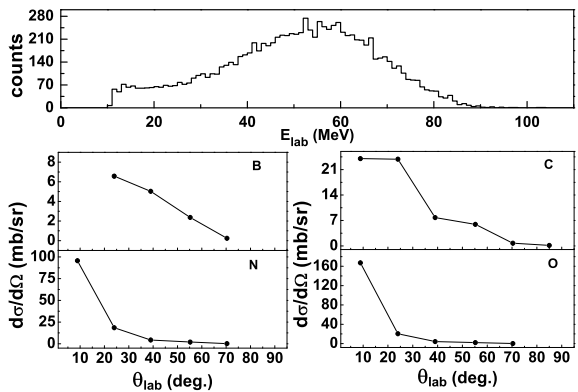}
\caption{Top: Energy spectrum of the N fragments at
$\theta_{lab}=31^\circ$. Bottom: Angular distributions of the B,
C, N and O fragments
(see text).}\label{fig1}
\end{figure}

\begin{figure}[!htbp]
\includegraphics[width=8cm]{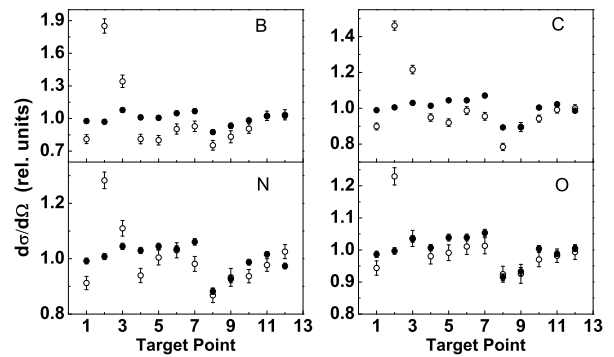}
\caption{Cross sections of the dissipative fragments B, C, N and O
at $\theta_{lab}=57^\circ$ (opened circles) and 31$^\circ$ (filled
circles) for the 12 different target spots (see text).}\label{fig2}
\end{figure}

In Fig. 2 we present the cross sections (in relative units) at the
12 different beam spots. These cross sections were obtained by
integrating all the counts of the energy spectra. The cross
sections were normalized as follows. For each reaction
fragment and angle, at a given beam spot, we divide the total number
of the counts by the solid angles of the telescopes and by
(M1+M2)/2, where M1 and M2 are the counts of the silicon
semiconductors. This allows us to scale out effects of variation
of the target thickness for different beam spots. Deviations of the ratio
M1/M2 from unity for different beam spots did not exceed 4$\%$.
Then, for each fragment and angle we obtain 12 numbers,
which are 12 cross sections (in relative units), corresponding to
the 12 beam spots. We scale these 12 numbers (with the
same scaling factor for given fragment and angle)
such that the average over the 12 cross sections is unity.
This procedure was applied for each fragment and angle. In
Fig. 2 the error bars are given by $\pm 1/N^{1/2}$, where $N$ is a
number of counts in the energy integrated spectrum, for each fragment and angle,
at a given beam spot. These numbers $N$ are generally different
for different fragments, angles and beam spots.

We also use another method to obtain the cross sections in Fig. 2.
For each reaction fragment and angle, at a given beam spot, we
divide the total number of the counts by the target thickness at
this beam spot, by 
Faraday cup charge counts and by the solid angles of the
telescopes. Again,
 for each fragment and angle we obtained
12 numbers, which are 12 cross sections (in relative units),
corresponding to the 12 beam spots. We scale these 12 numbers to
have the average over the 12 beam spots equal to unity, for given
fragment and angle. This procedure was applied for each
fragment at each angle. We found that both methods produce the
data which agree within the statistical accuracy. 
The cross sections in Fig. 2 are clearly not the same for
different beam spots. To quantify the cross section variations, in
Fig. 3 we plot the probability distributions of absolute values of the
deviations of the cross sections for each fragment and angle, at
different beam spots, from the average value of unity.
 The deviations are scaled with the
$1/N^{1/2}$ factors, where $N$ is a number of counts in the whole
spectrum for given fragment and angle at each beam spot. If
these deviations originated from a finite number of the counts the
probability distributions would be Gaussian with a standard
deviation of unity. However the data are scattered on a much wider
interval. Indeed, the probability for deviations greater than
three standard deviations is 50$\%$ for 5 cases in Fig. 3, and
33.33$\%$, 25$\%$, 8.33$\%$ for the rest 3 cases. The
corresponding probability for the Gaussian distribution is
0.27$\%$.

\begin{figure}[!htbp]
\includegraphics[width=8cm]{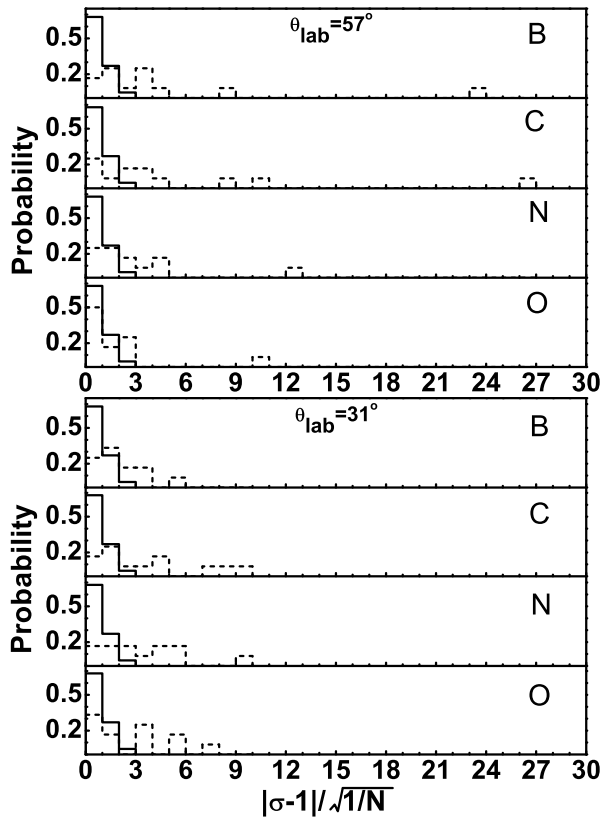} \caption{Dotted histograms: Probability
distributions of absolute values of the measured relative cross
sections deviations from averaged value of unity for the B, C, N
and O fragments at $\theta_{lab}=57^\circ$ and 31$^\circ$. Solid
histogram: Gaussian distributions with a standard deviation of
unity (see text).}\label{fig3}
\end{figure}

\begin{figure}[!htbp]
\includegraphics[width=8cm]{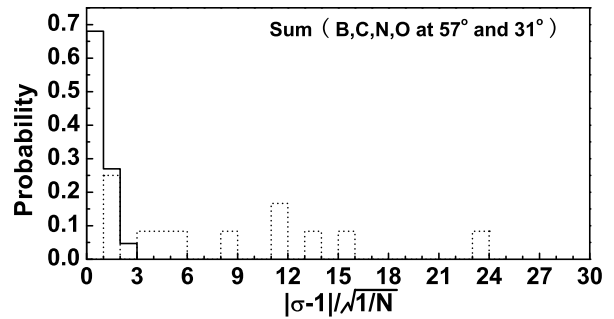} \caption{Dotted histogram: Probability
distribution of absolute values of deviations of the measured
 summed cross sections from averaged value of unity. Solid
histogram: Gaussian distribution with a standard deviation of
unity (see text).}\label{fig4}
\end{figure}

In Fig. 4 we plot the probability distribution for the deviations
from the average unity of the cross sections summed over all the
fragments and angles. For each beam spot, we sum 8 previously
obtained partial cross sections (in the
 relative units)
to obtain the summed cross section. We apply this procedure for
each of the 12 beam spots. As a result we obtain 12 numbers which
are the summed cross sections in relative units. We normalize
these cross sections such that the cross section averaged over all
the 12 beam spots is unity. In Fig. 4, absolute values of the
deviations of the cross sections, measured at different beam
spots, from the averaged value of unity are scaled with the
$1/N^{1/2}$ factors, where $N$ is a sum of the counts numbers for
all the fragments and angles at a given beam spot.
 In this way, for each of the 12 beam spots,
we increase the total number of counts resulting in a much better
statistics of the data
 as compared with the analysis in Fig. 3. In Fig. 4 we also plot Gaussian
distribution with a standard deviation of unity. This Gaussian
distribution is expected if finite number of counts would be the
only source for the deviations.
 We see a dramatic spread of the data in Fig. 4 as compared with the Gaussian
distribution. Indeed, the
probability for deviations more than four standard deviations is
67$\%$, i.e. four orders of magnitude more than the value of
0.006$\%$ expected from the Gaussian statistics with a standard
deviation of unity. We conclude that the experimental data do
indicate a statistically significant dramatic
non-reproducibility of the cross sections for different beam
spots. In some cases
 the measured cross sections differ up to a factor of 1.5-2 for different beam spots
(see Fig. 2).

Notice that visible correlations, for each individual angle,
between the cross sections corresponding to the different
fragments is consistent with the interpretation~\cite{kunPRL}:
Though the B, C, N and O cross sections summed over very large
number of exit channels have different random components for
different beam spots these cross sections set up in a correlated
manner for a given angle.
A smaller amplitude of the cross section variations at
$\theta_{lab}=31^\circ$ as compared to that for
$\theta_{lab}=57^\circ$ is explained by the bigger relative
contribution of direct-like processes for forward angles. 
This is because cross sections of the direct-like processes are stable
with respect to very small perturbations.
For $\theta_{lab}=57^\circ$, the decrease of the amplitude of the
cross sections variations with increase of the difference in mass
numbers between the ejectiles and the projectile F is because the
smaller this difference is the more are the relative contributions of
direct-like processes.

The interpretation of the obtained results is based on a picture
of formation of the IC with strongly overlapping resonances
and a very slow phase relaxation between the coherently excited states with
different total spin values~\cite{kunPRL}.
The intrinsic excitation energy of the IC,
$E^*\simeq 20$ MeV, is obtained by subtracting the deformation
energy, $E_{def}\simeq 20$ MeV, and the rotational energy for the spin
values close to grazing orbital momentum, $E_{rot}\simeq 55$ MeV~\cite{RomCat}, 
from the total excitation energy $E=95.14$ MeV. For $E^*\simeq
20$ MeV, $D\simeq 10^{-7}$ MeV. We evaluate the effective
strength, $|K|$, of the ``target-environmental'' perturbations to
be of order of the atomic electron effects in DHIC~\cite{kunPRL}.
These perturbations are due to the different distributions of
electro-magnetic fields, defects, etc. for the different beam spots.
 From~\cite{kunPRL}, we obtain $K\sim \pm
10^{-6}D\simeq\pm 10^{-7}$ eV, which about 13 orders of magnitude
less than the effective nucleon-nucleon interaction! Such small perturbations
may be considered as nominally ``infinitesimally small''.
Yet, it was suggested~\cite{kunPRL} 
that even such small perturbations may strongly
affect the cross sections. The basic point of the interpretation~\cite{kunPRL} 
is that the cross sections summed over very large number of exit
channels are determined by the values and signs of the quantities
$N_H^{1/2}<\phi_\mu^J|\phi_\nu^I>$, where $N_H\to\infty$ is a
dimension of Hilbert space and $<\phi_\mu^J|\phi_\nu^I>$ are
scalar products of the many-body resonance eigenstates
$\phi_\mu^J$, $\phi_\nu^I$ with different total spins $J\neq I$
and $\mu,\nu$ being running indices. The  $\phi_\mu^J$ and
$\phi_\nu^I$ are orthogonal, i.e. $<\phi_\mu^J|\phi_\nu^I>=0$.
Then, applying a proper procedure, in the limit $N_H\to\infty$, the quantities
$N_H^{1/2}<\phi_\mu^J|\phi_\nu^I>$ are uncertainties~\cite{App1}. The
anomalous sensitivity of the cross sections occurs provided these
uncertainties do not vanish. But then the values, in
particular, the signs of the $N_H^{1/2}<\phi_\mu^J|\phi_\nu^I>$
are set at random implying that the cross sections are
unpredictable~\cite{App2}. This also demonstrates that an
infinitesimally small perturbation, which leads to infinitesimally
small changes of the $\phi_\mu^J$ and $\phi_\nu^I$, can change the
signs of the uncertainties thereby resulting in a very large
change of the cross section~\cite{kunPRL}. If so, the effect
indicates deterministic randomness in complex quantum collisions
which occurs due to the instability with respect to
infinitesimally small perturbations.

Usually, an effect is considered to be reliably established if it
can be reproduced in independent measurements with nominally identical
experimental conditions. We realize that the
experimental data presented in this Letter are in contradiction
with this conventional point of view. Instead, we deal with 
``reproducible non-reproducibility'' for the different nominally identical experiments.
Namely, the effect can be considered as a real one
provided the non-reproducibility of the cross sections 
will be confirmed in new independent experiments performed at other
tandem accelerators. No matter how the
 interpretation~\cite{kunPRL} may be viewed, the importance of
the presented {\sl experimental results} is far beyond the nuclear
physics field calling for a new line of thinking. 
In view of the fundamental role of deterministic
randomness in classical physics for the understanding of
statistical laws, in particular statistical relaxation, from dynamics~\cite{CasChir},
 it is highly desirable to test a
possible deterministic randomness in quantum many-body systems in
new experiments. 
``No matter how negligible the probability, in view of the thousands of
experiments already done, any new possibility like quantum chaos
should be used carefully to check the fundamental equations in the
laboratory again and again''~\cite{CasChir}. Our
results do present a real possibility for
the new and unexpected phenomenon of instability of the cross sections
is complex quantum collisions.

\begin{acknowledgments}
This work has been supported by the National Natural Science Foundation of China
(Grants No. 10475101 and No. 10675149), by the Chinese Academy of Sciences (Grant
No. 0730030YF0) and by Conacyt (Grant No. 43375).
\end{acknowledgments}


\begin{thebibliography}{99}

\bibitem{CasChir} G. Casati and B.V. Chirikov, in {\sl Quantum Chaos: Between Order
and Disorder}, edited by G. Casati and B.V. Chirikov (Cambridge
University Press, 1995), pp. 3-53, and references therein.

\bibitem{kunPRL} S.Yu. Kun, Phys. Rev. Lett. {\bf 84}, 423 (2000),  and references therein.

\bibitem{App3} Advantage of collision experiments is that these allow to obtain
information about IC without perturbing it thereby avoiding a problem of measurement
in quantum mechanics~\cite{CasChir}.

\bibitem{kunAIP}
S.Yu. Kun, in {\sl Non-Equilibrium and Nonlinear Dynamics in
Nuclear and Other Finite Systems}, edited by Zhuxia Li, Ke Wu,
Xizhen Wu, Enguang Zhao, and Fumihiko Sakata, AIP Conf. Proc. No.
597 (AIP, Melville, NY, 2001), p. 319. (2001).

\bibitem{WQijmpe} Wang Qi et al., Int. J. Mod. Phys. E {\bf 12},
377 (2003).

\bibitem{WQref2} J.F. Zegler, {\sl Helium Stopping Powers and Ranges in all Elements},
Vol. 4 (Pergamon Press, NY, 1977), p. 45.

\bibitem{WQref1} Xu Guoji et al., At. En. Sci. Tech. {\bf 25}, 34 (1991).

\bibitem{RomCat} I. Berceanu et al.,
 Phys. Rev. C {\bf 57}, 2359 (1998).

\bibitem{App1} A possible way to quantify the uncertainties is to carry integration
in the scalar products over the whole integration volume except
an arbitrary located infinitesimally small element of this volume
$\Delta V$. Then, the properly taken limits $N_H\to\infty$,
$\Delta V\propto 1/N_H^{1/2}\to 0$ lead to (i) recovery of the
scalar products and, (ii) non-vanishing of the
uncertainties (S.Yu. Kun, work in progress).

\bibitem{App2} This is especially clear for digital computer calculations for which
the $<\phi_\mu^J|\phi_\nu^I>$ do not vanish exactly due to the
finite accuracy. Then, the signs
of the $N_H^{1/2}<\phi_\mu^J|\phi_\nu^I>$ are unpredictable
in principle. A similar argument is often used to
illustrate the impossibility of predicting long-time evolution of
classical chaotic systems due to the unavoidable computational errors~\cite{CasChir}.

\end{thebibliography}
\end{document}